\documentclass[prd,twocolumn,superscriptaddress,showpacs,nofootinbib,preprintnumbers]{revtex4}

\usepackage{amsmath}
\usepackage{amsfonts}
\usepackage{graphicx}
\usepackage{dcolumn}
\usepackage{hyperref}

\begin{document}
\tolerance=5000
\def\pp{{\, \mid \hskip -1.5mm =}}
\def\cL{{\cal L}}
\def\be{\begin{equation}}
\def\ee{\end{equation}}
\def\bea{\begin{eqnarray}}
\def\eea{\end{eqnarray}}
\def\tr{{\rm tr}\, }
\def\nn{\nonumber \\}
\def\e{{\rm e}}

\title{Inhomogeneous Equation of State of the Universe: Phantom Era, Future
Singularity and Crossing the Phantom Barrier}

\author{Shin'ichi Nojiri}
\affiliation{Department of Applied Physics,
National Defence Academy, Hashirimizu Yokosuka 239-8686, Japan}
\author{Sergei D.~Odintsov}
\affiliation{Instituci\`o Catalana de Recerca i Estudis
Avan\c{c}ats (ICREA)  and Institut d'Estudis Espacials de Catalunya (IEEC),
Edifici Nexus, Gran Capit\`a 2-4, 08034 Barcelona, Spain}

\date{\today}

\vskip 1pc
\begin{abstract}
The dark energy universe equation of state (EOS) with inhomogeneous,
Hubble parameter dependent term is considered. The motivation to introduce
such a term comes from time-dependent viscosity considerations and 
modifications of general relativity. For several explicit examples of such 
EOS it is demonstrated how the type of future singularity changes, how the 
phantom epoch emerges and how crossing of phantom barrier occurs.
Similar cosmological regimes are considered for the universe with two 
interacting fluids and for universe with implicit EOS. For instance, the 
crossing of 
phantom barrier is realized in easier way, thanks to the presence of
inhomogeneous term. The thermodynamical dark energy model is presented 
where the universe entropy may be positive even at phantom era as a result
of crossing of $w=-1$ barrier.

\end{abstract}


\maketitle
\vskip 1pc

\section{Introduction}

The increasing number of evidences from the observational data indicates
that current universe lives in a narrow strip near $w=-1$ (where w is
the equation of state (EOS) parameter), quite probably being below $-1$
in so-called phantom region.
It is also assumed that modern universe is filled with some mysterious,
negative pressure fluid (dark energy) which represents about 70 percents of 
total
energy in the universe. (The simplest, phenomenological approach is
to consider that this fluid satisfies to EOS with constant $w$). The
origin of this dark energy is really dark: the proposed explanations
vary from the modifications of gravity to the introduction of new fields
(scalars, spinors, etc) with really strange properties. Moreover,
forgetting for the moment about the origin of dark energy, even
more-less satisfactory mechanism of evolving dark energy is missing, so
far. At best, each of existing theoretical models for dark energy
explains some specific element(s) of late-time evolution, lacking the
complete understanding. Definitely, the situation may be improved with the
new generation of observational data when they will present the realistic
evolving EOS
of dark energy during sufficiently large period.

The most strange (if realistic) era in the universe evolution is phantom
era. There are many attempts to describe the phantom cosmology (see, for
instance, \cite{CKW,E} and references therein), especially near to
future, finite-time singularity (Big Rip) which is the essential element
of classical phantom cosmology. (Note that quantum effects may basically
provide the escape from future, finite type singularity, for recent
discussion, see\cite{final,NOT}).  Unfortunately, the easiest way
to describe the phantom cosmology in the Lagrangian formulation leads to
the necessity of the introduction of not very appreciated scalar with
negative kinetic
energy\cite{caldwell}. Another, easy way is to use some phenomenological
EOS which may produce dark epoch of the universe (whatever it is).
It is remarkable that such description shows the possibility of other
types
of future, finite type singularity. For instance, even when EOS is
suddenly phantomic (near to rip time where negative pressure diverges),
the sudden singularity occurs
\cite{Barrow}. There may exist future singularities where energy/pressure
is
finite
at rip time, for classification of future singularities,
  see \cite{NOT}. They may occur even in modified gravity at late
times,
see\cite{gravity} for explicit examples. Nevertheless, it is remarkable
that effective phantom phase may be produced also in string-inspired
gravities\cite{string}.

The present paper is devoted to study the phantom cosmology and related
regimes (for instance, crossing of phantom divide) when phenomenological
equation of state of the universe is inhomogeneous. In other words,
it contains terms dependent explicitly from Hubble parameter (or, even
from its derivatives). Definitely, one needs quite strong motivation
for such modification of dark energy EOS. The first one comes from the
consideration of time-dependent bulk viscosity\cite{BG,viscosity}.
(For earlier discussion of cosmology with time-dependent bulk viscosity, 
see see also \cite{Barrow2}.)
Actually, it was constructed the specific model of dark energy with
possibility of crossing of phantom divide due to time-dependent bulk
viscosity \cite{BG}. The construction of EOS from symmetry considerations
\cite{symmetry} indicates to the necessity of some inhomogeneous
correction. Finally, big number of gravities: from low-energy string
effective actions to gravity with higher derivative terms or with inverse
terms on curvatures modifies the FRW equations in requested form.

The paper is organized as follows. In the next section we consider
spatially-flat FRW universe filled by the ideal fluid with specific, dark
energy EOS
\cite{final}. Short review of four types of future singularity for
different choices of EOS parameters is given, following to ref.\cite{NOT}.
The inhomogeneous term of specific form is introduced to EOS.
The role of such term in the transition of different types of singularity
to another ones is investigated. The cosmological regimes crossing phantom
barrier due to such  terms are explicitly constructed.
Finally, the dependence of the inhomogeneous term from Hubble parameter
derivatives is briefly discussed as well as emerging oscillating universe.
Section three is devoted to the study of similar questions when FRW
universe is filled by the interacting mixture of two fluids. The
modification of two fluids EOS by inhomogeneous term is again considered.
The explicit example of late-time cosmology (which may be oscillating one)
quite naturally crossing
the phantom divide in such a universe is presented. It is interesting
that inhomogeneous term may effectively compensate the interaction between
two fluids. In the section four we discuss the FRW cosmology admitting the
crossing of barrier $w=-1$ due to specific form of the implicit dark
energy EOS proposed in ref.\cite{Stf}. Again, the generalized, Hubble
parameter dependent EOS is considered. Some thermodynamical dark energy
model passing the barrier $w=-1$ is constructed, based on above EOS.
It is demonstrated that in such a model the universe entropy may be
positive even during the phantom era.
Some summary and outlook are given in the discussion section.
The Appendix deals with couple simple versions of modified gravity
which may predict the requested generalization of EOS.

\section{FRW cosmology with inhomogeneous dark energy equation of state}

In the present section we make brief review of FRW cosmology with explicit dark
energy equation of state (power law). The modification of EOS by Hubble
dependent term (constrained by energy conservation law) is done and its
role to FRW cosmology evolution is investigated.
The starting FRW universe metric is:
\be
\label{FRW}
ds^2=-dt^2 + a(t)^2 \sum_{i=1}^3  \left(dx^i\right)^2\ .
\ee
In the FRW universe, the energy conservation law can be expressed as
\be
\label{ppH1}
0=\dot\rho + 3H\left(p + \rho\right)\ .
\ee
Here $\rho$ is energy density, $p$ is pressure. The Hubble rate $H$ is defined
by $H\equiv \dot a/a$. When $\rho$ and $p$ satisfy the following simple EOS:
\be
\label{ppH2}
p=w\rho\ ,
\ee
and if $w$ is a constant, Eq.(\ref{ppH1}) can be easily integrated:
\be
\label{ppH3}
\rho=\rho_0 a^{-3(1+w)}\ .
\ee
Using the first  FRW equation
\be
\label{pH3}
\frac{3}{\kappa^2}H^2=\rho\ ,
\ee
the well-known solution follows
\be
\label{ppH4}
a=a_0 \left(t - t_1\right)^\frac{2}{3(w+1)}\quad \mbox{or}
\quad a_0 \left(t_2 - t\right)^\frac{2}{3(w+1)}\ ,
\ee
when $w\neq -1$, and
\be
\label{ppH5}
a=a_0\e^{\kappa t\sqrt{\frac{\rho_0}{3}}}
\ee
when $w=-1$.
In (\ref{ppH4}), $t_1$ and $t_2$ are constants of the integration.
Eq.(\ref{ppH5}) expresses the deSitter universe.
In (\ref{ppH4}), since the exponent $2/3(w+1)$ is not integer in general, we
find $t>t_1$ or $t<t_2$
so that $a$ should be real number.
If the exponent $2/3(w+1)$ is positive, the first solution in (\ref{ppH4})
expresses the expanding
universe but the second one expresses the shrinking universe.
If the exponent $2/3(w+1)$ is negative, the first solution in (\ref{ppH4})
expresses the shrinking
universe but the second one expresses the expanding universe.
In the following, we only consider the case that the universe is expanding.
Then for the second solution, however, there appears a singularity in a finite
time at $t=t_2$,
which is called the Big Rip singularity ( for discussion of
phantom cosmology near Big Rip and related questions, see\cite{CKW,E}
and references therein) when
\be
\label{ppH5b}
w<-1\ .
\ee


In general, the singularities may behave in different ways.
One may classify the future singularities
as following\cite{NOT}:
\begin{itemize}
\item  Type I (``Big Rip'') : For $t \to t_s$, $a \to \infty$,
$\rho \to \infty$ and $|p| \to \infty$
\item  Type II (``sudden'') : For $t \to t_s$, $a \to a_s$,
$\rho \to \rho_s$ or $0$ and $|p| \to \infty$
\item  Type III : For $t \to t_s$, $a \to a_s$,
$\rho \to \infty$ and $|p| \to \infty$
\item  Type IV : For $t \to t_s$, $a \to a_s$,
$\rho \to 0$, $|p| \to 0$ and higher derivatives of $H$ diverge.
This also includes the case when $\rho$ ($p$) or both of them
tend to some finite values while higher derivatives of $H$ diverge.
\end{itemize}
Here $t_s$, $a_s$ and $\rho_s$ are constants with $a_s\neq 0$.
The type I may correspond to the Big Rip singularity \cite{CKW},
which emerges when $w<-1$ in (\ref{ppH2}).
The type II corresponds to the sudden future
singularity  \cite{Barrow} at which
$a$ and $\rho$ are finite but $p$ diverges.
The type III appears for the model with $p=-\rho-A \rho^{\alpha}$
\cite{Stefancic}, which is
different from the sudden future singularity
in the sense that $\rho$ diverges.
This type of singularity has been discovered in the model
of Ref.~\cite{final} where the corresponding Lagrangian
model of a scalar field with potential has been constructed.

One may start from the dark energy EOS as
\be
\label{EOS1}
p=-\rho - f(\rho)\ ,
\ee
where $f(\rho)$ can be an arbitrary function in general.
The function $f(\rho)\propto \rho^\alpha$ with a constant $\alpha$
was proposed in Ref. \cite{final} and was investigated in detail in
Ref.\cite{Stefancic}.
Using (\ref{ppH1}) for such choice,  the scale factor is given by
\be
\label{EOS4}
a=a_0\exp\left(\frac{1}{3} \int \frac{d\rho}{f(\rho )} \right)\ .
\ee
Using (\ref{pH3}) the cosmological time may be found
\be
\label{tint}
t=\int \frac{d \rho}{\kappa \sqrt{3\rho} f(\rho)}\ ,
\ee
In case
\be
\label{ppH6}
f(\rho)=A\rho^\alpha\ ,
\ee
by using Eq.(\ref{EOS4}), it follows
\be
\label{EOS26}
a = a_0 \exp \left[\frac{\rho^{1-\alpha}}{3(1-\alpha)A} \right]\ .
\ee
When $\alpha> 1$, the scale factor remains finite even if $\rho$ goes to
infinity.
When $\alpha<1$,  $a\to \infty$ $(a\to 0)$ as $\rho\to \infty$ for $A>0$
$(A<0)$.
Since the pressure is now given by
\be
\label{EOS27}
p= -\rho - A\rho^\alpha\ ,
\ee
$p$ always diverges when $\rho$ becomes infinite.
If $\alpha>1$, the EOS parameter $w=p/\rho$
also goes to infinity, that is,
$w \to +\infty$ ($-\infty$) for $A<0$ $(A>0$).
When $\alpha<1$, we have $w\to -1+0$ ($-1-0$)
for $A<0$ $(A>0$) as $\rho \to \infty$.

By using Eq.(\ref{tint}) for (\ref{ppH6}), one finds\cite{NOT}
\be
\label{EOS31}
t= t_0 + \frac{2}{\sqrt{3}\kappa A}\frac{\rho^{-\alpha+1/2}}{1-2\alpha}\ ,\quad
{\rm for} \quad \alpha \neq \frac12\ ,
\ee
and
\be
\label{EOS32}
t= t_0 + \frac{{\rm ln}\left(\frac{\rho}{\rho_0}\right)}{\sqrt{3}\kappa A}\ ,
\quad {\rm for}\quad \alpha=\frac12\,.
\ee
Therefore if $\alpha\leq 1/2$, $\rho$ diverges in an infinite future or past.
On the other hand, if $\alpha>1/2$, the divergence of $\rho$
corresponds to a finite future or past.
In case of finite future, the singularity could be regarded as a Big Rip or
type I singularity.

For the choice (\ref{ppH6}),  the following cases were discussed \cite{NOT}:
\begin{itemize}
\item In case $\alpha=1/2$ or $\alpha=0$, there does not appear any
singularity.
\item In case $\alpha>1$, Eq.(\ref{EOS31}) tells that when $t\to t_0$, the
energy density behaves
as $\rho\to\infty$ and therefore $|p|\to \infty$ due to (\ref{EOS27}).
Eq.(\ref{EOS26}) shows
that the scale factor $a$ is finite even if $\rho\to \infty$. Therefore
$\alpha>1$ case
corresponds to type III singularity.
\item $\alpha=1$ case corresponds to the case (\ref{ppH2}) if we replace $-1 -
A$ with $w$.
Therefore if $A>0$, there occurs the Big Rip or type I singularity but if
$A\leq 0$,
there does not appear future singularity.
\item In case $1/2<\alpha<1$, when $t\to t_0$, all of $\rho$, $|p|$, and $a$
diverge if $A>0$
then this corresponds to type I singularity.
\item In case $0<\alpha<1/2$, when $t\to t_0$, we find $\rho$, $|p|\to 0$ and
$a\to a_0$ but by
combining (\ref{EOS26}) and (\ref{EOS31}), we find
\be
\label{typeIV}
\ln a \sim \left|t-t_0\right|^{\frac{\alpha-1}{\alpha - 1/2}}\ .
\ee
Since the exponent $(\alpha -1)/(\alpha - 1/2)$ is not always an integer, even
if $a$
is finite, the higher derivatives of $H$ diverge in general. Therefore this
case corresponds to type IV singularity.
\item In case $\alpha<0$, when $t\to t_0$, we find $\rho\to 0$, $a\to a_0$ but
$|p|\to \infty$.
Therefore this case corresponds to type II singularity.
\end{itemize}
Hence, the brief review of FRW cosmology with specific homogeneous EOS as
well
as its late-time behaviour (singularities) is given (see \cite{NOT} for more
detail).

At the next step, we will consider the inhomogeneous EOS for dark energy,
so that the dependence from Hubble parameter is included in EOS.
The motivation for such EOS comes from including of time-dependent bulk
viscosity in ideal fluid EOS \cite{BG} or from the modification of gravity
(see Appendix).
Hence, we suggest the following EOS
\be
\label{pH1}
p=-\rho + f(\rho) + G(H)\ .
\ee
where $G(H)$ is some function.
Then the energy conservation law (\ref{ppH1}) has the following form:
\be
\label{pH2}
0= \dot \rho + 3H\left(f(\rho) + G(H)\right)\ .
\ee
By using the first FRW equation (\ref{pH3}) and assuming the expanding universe
$\left(H \geq 0\right)$, one finds
\be
\label{pH4}
\dot\rho = F ( \rho ) \equiv -3\kappa\sqrt{\frac{\rho}{3}}\left( f(\rho)
+ G\left(\kappa\sqrt{\rho/3}\right)\right)\ .
\ee
or
\be
\label{pH5o}
G(H)=- f\left(3H^2/\kappa^2\right) + \frac{2}{\kappa^2}\dot H\ .
\ee
Hence, one  can express $G(H)$ in terms of $f$ as above.

As a first example, let assume that EOS(\ref{ppH2}) could be modified as
\be
\label{ppH7}
p=w_0\rho + w_1 H^2\ .
\ee
Using (\ref{pH3}), it follows
\be
\label{ppH8}
p=\left(w_0 + \frac{\kappa^2 w_1}{3}\right)\rho\ .
\ee
Therefore $w$ is effectively shifted as
\be
\label{ppH8b}
w\to w_{\rm eff}\equiv w_0 + \frac{\kappa^2 w_1}{3}\ .
\ee
Then even if $w_0<-1$, as long as $w_{\rm eff}>-1$, there does not occur
the Big Rip singularity. From another side one can start with quintessence
value of $w_0$, the inhomogeneous EOS (\ref{ppH8}) with sufficiently
negative $w_1$ brings the cosmology to phantom era.

As a second example, we assume  $f(\rho)$  (\ref{ppH6}) is modified as
\be
\label{ppH9}
f(\rho)=A\rho^\alpha \to f(\rho) + G(H) = - A\rho^{\alpha} - B H^{2\beta}\ .
\ee
By using the first FRW equation (\ref{pH3}),
we find  $f(\rho)$ is modified as
\bea
\label{pppH1}
&& f_{\rm eff}(\rho)
= f(\rho) + G(H) = - A\rho^{\alpha} - B' \rho^\beta\ ,\nn
&& B'\equiv B \left(\frac{\kappa^2}{3}\right)^\beta\ .
\eea
If $\beta>\alpha$, when $\rho$ is large,  the second term in (\ref{pppH1})
becomes dominant:
\be
\label{ppH10}
f_{\rm eff}(\rho) \to B' \rho^\beta\ .
\ee
On the other hand, if $\beta<\alpha$, the second term becomes dominant and we
obtain (\ref{ppH10})
again when $\rho\to 0$.
In case of (\ref{ppH6}) without $G(H)$, when $1/2<\alpha<1$, there is the type
I singularity
where $\rho$ goes to infinity in a finite time. When  $G(H)$ is given by
(\ref{ppH9}),
if $\beta>\alpha$, the second term in (\ref{pppH1}) becomes dominant and
therefore if $\beta>1$,
instead of type I singularity there occurs type III singularity.
In case of (\ref{ppH6}) with $\alpha>1$, the type III singularity appears
before  $G(H)$ is included.
Even if we include $G(H)$ with $\beta>\alpha>1$, we obtain the type III
singularity again and the
structure of the singularity is not changed qualitatively.
For (\ref{ppH7}) without $G(H)$, when $0<\alpha<1/2$ or $\alpha<0$, there
appears the type IV or
type II singularity where $\rho$ tends to zero. Since the second term becomes
dominant
if $\beta<\alpha$, if $\beta<0$, the type IV singularity for $0<\alpha<1/2$
case becomes the
type II singularity but the type II singularity for $\alpha<0$ is not
qualitatively changed.

In accordance with the  previous cases, one finds
\begin{itemize}
\item In case $\alpha>1$, for most values of $\beta$, there occurs type III
singularity.
In addition to the type III singularity, when $0<\beta<1/2$, there  occurs type
IV
singularity and when $\beta<0$, there  occurs type II singularity.
\item $\alpha=1$ case, if $\beta>1$, the singularity becomes type III.
$\beta=1$ case corresponds to (\ref{ppH7}). If $\beta<1$ and $A>0$, there
occurs the Big Rip
or type I singularity. In addition to the type I singularity, we  have type IV
singularity
when $0<\beta<1/2$ and type II when $\beta<1$.
\item In case $1/2<\alpha<1$, one sees singularity of type III for $\beta>1$,
type I for $1/2\leq\beta<1$
(even for $\beta=1/2$) or $\beta=1$ and $B'>0$ ($B>0$)  case. In addition to
type I, type IV case occurs for $0<\beta<1/2$, and type II for $\beta<0$.
\item In case $\alpha=1/2$, we have singularity of type III for $\beta>1$, type
I for $1/2<\beta<1$
or $\beta=1$ and $B'>0$ ($B>0$), type IV for $0<\beta<1/2$, and type II for
$\beta<0$.
When $\beta=1/2$ or $\beta=0$, there does not appear any singularity.
\item In case $0<\alpha<1/2$, we find type IV for $0<\beta<1/2$, and type II
for $\beta<0$.
In addition to type IV singularity,  there occurs singularity of
type III for $\beta>1$, type I for $1/2\leq\beta<1$ or $\beta=1$ and $B'>0$
($B>0$)  case.
\item In case $\alpha<0$, there will always occur type II singularity.
In addition to type II singularity, we  have a singularity of
type III for $\beta>1$, type I for $1/2\leq\beta<1$ or $\beta=1$ and $B'>0$
($B>0$)  case.
\end{itemize}

Thus, we demonstrated how the modification of EOS by Hubble dependent,
inhomogeneous term changes the structure of singularity in late-time dark
energy universe.

We now consider general case and assume $F(\rho)$ in (\ref{pH4}) behaves as
\be
\label{pH5}
F(\rho)\sim F_0 \rho^\alpha\ ,
\ee
with constant $F_0$ and $\alpha$ in a proper limit (e.g. for large $\rho$ or
small $\rho$).
Then when $\alpha\neq 1$, Eq.(\ref{pH4}) can be integrated as
\be
\label{pH6}
F_0\left(t - t_c\right) \sim \frac{\rho^{1-\alpha}}{1-\alpha}\ ,
\ee
that is,
\be
\label{pH7}
\rho \sim \left( (1-\alpha)F_0\left(t - t_c\right)
\right)^{\frac{1}{1-\alpha}}\ .
\ee
Here $t_c$ is a constant of the integration.
When $\alpha=1$, the energy becomes
\be
\label{pH8}
\rho = \rho_0\e^{F_0 t}\ ,
\ee
with a constant of integration $\rho_0$.
By using the first FRW equation (\ref{pH3}), the scale factor may be found
\be
\label{pH9}
a=a_0 \e^{\pm \frac{2\kappa}{(3-2\alpha)\sqrt{3}F_0}
\left( (1-\alpha)F_0\left(t - t_c\right) \right)^{\frac{3-2\alpha
}{2(1-\alpha)}}}\ ,
\ee
when $\alpha \neq 1$ and
\be
\label{pH10}
a=a_0 \e^{\frac{2\kappa}{F_0}\sqrt{\frac{\rho_0}{3}}\e^{\frac{F_0 t}{2}}}\ ,
\ee
when $\alpha=1$.

In \cite{NOT}, there has been given an explicit example of the EOS where
crossing of $w=-1$ phantom divide occurs:
\be
\label{EOS7}
a(t)=a_0 \left(\frac{t}{t_s - t}\right)^n \ .
\ee
Here $n$ is a positive constant and  $0<t<t_s$.
The scale factor diverges in a finite time ($t \to t_s$)
as in the Big Rip.
Therefore $t_s$ corresponds to the life time of the universe.
When $t \ll t_s$, $a(t)$ evolves as $t^n$,
which means that the effective EOS
is given by $w=-1 + 2/(3n)>-1$.
On the other hand, when $t \sim t_s$, it appears $w=-1 - 2/(3n)<-1$.
The solution  (\ref{EOS7}) has been obtained with
\be
\label{EOS14}
f(\rho) = \pm \frac{2\rho}{3n}\left\{ 1
   - \frac{4n}{t_s}\left(\frac{3}{\kappa^2\rho}
\right)^{\frac{1}{2}}\right\}^{\frac{1}{2}}\ .
\ee
Therefore the EOS needs to be double-valued in order for
the transition to occur between the region $w<-1$ and the region $w>-1$.
Then in general, there could not be one-to-one correspondence between $p$ and
$\rho$
in the above EOS.
In such a case, instead of (\ref{pH1}), we may suggest the implicit,
inhomogeneous equation of the state
\be
\label{pH11}
F(p,\rho,H)=0\ .
\ee
The following example may be of interest:
\be
\label{pH12}
\left(p+\rho\right)^2 - C_0 \rho^2 \left(1 - \frac{H_0}{H}\right)=0\ .
\ee
Here $C_0$ and $H_0$ are positive constants. Combining
(\ref{pH12}) with
the energy conservation law
(\ref{pH2}) and the first FRW equation (\ref{pH3}), one can delete $p$ and
$\rho$ as
\be
\label{pH13}
{\dot H}^2=\frac{9}{4}C_0 H^4 \left(1 - \frac{H_0}{H}\right)\ ,
\ee
which can be integrated as
\be
\label{pH14}
H=\frac{16}{9C_0^2 H_0 \left(t-t_-\right)\left(t_+ - t\right)}\ .
\ee
Here
\be
\label{pH15}
t_\pm = t_0 \pm \frac{4}{3C_0 H_0}\ ,
\ee
and $t_0$ is a constant of the integration. Hence
\bea
\label{pH16}
p&=&-\rho\left\{1+\frac{3C_0^2}{4H_0}\left(t-t_0\right)\right\}\ ,\nn
\rho &=&\frac{2^8}{3^3C_0^4 H_0^2 \kappa^2 \left(t-t_-\right)^2
\left(t_+ - t\right)^2}\ .
\eea
In (\ref{pH14}), since $t_-<t_0<t_+$, as long as $t_-<t<t_+$, the Hubble rate
$H$ is positive.
The Hubble rate $H$ has a minimum $H=H_0$ when $t=t_0=\left(t_- + t_+\right)/2$
and diverges when $t\to t_\pm$.
Then we may regard $t\to t_-$ as a Big Bang singularity and $t\to t_+$ as a Big
Rip one.
As clear from (\ref{pH16}), the parameter $w=p/\rho$ is larger than $-1$ when
$t_-<t<t_0$
and smaller than $-1$ when $t_0<t<t_+$. Therefore there occurs crossing of
phantom divide $w=-1$
when $t=t_0$ thanks to the effect of inhomogeneous term in EOS.

One more example may be of interest:
\be
\label{pH17}
\left(\rho + p\right)^2 + \frac{16}{\kappa^4 t_0^2}\left(h_0 - H\right)
\ln\left(\frac{h_0 - H}{h_1} \right)=0\ .
\ee
Here $t_0$, $h_0$, $h_1$ are constants and $h_0>h_1>0$.
A solution is given by
\bea
\label{pH18}
&& H=h_1 - h_1\e^{-t^2/t_0^2}\ ,\quad
\rho = \frac{3}{\kappa^2}\left(h_1 - h_1\e^{-t^2/t_0^2}\right)^2\ ,\nn
&& p= -\frac{3}{\kappa^2}\left(h_1 - h_1\e^{-t^2/t_0^2}\right)^2
      - \frac{4h_1 t}{\kappa^2 t_0^2}\e^{-t^2/t_0^2}\ .
\eea
Hence,
\be
\label{pH19}
\dot H= \frac{2h_1 t}{t_0^2}\e^{-t^2/t_0^2}\ .
\ee
Using the energy conservation law (\ref{pH2}) and the first FRW equation
(\ref{pH3}), the second FRW equation may be found:
\be
\label{pH20}
- \frac{2}{\kappa^2}\dot H = \rho + p \ .
\ee
As in (\ref{pH19}), $\dot H$ is negative when $t<0$ and positive when $t>0$.
Eq.(\ref{pH20}) tells that the effective parameter $w=p/\rho$ of the equation
of the state is
$w>-1$ when $t<0$ and $w<-1$ when $t>0$.
As we find the Hubble rate $H$ goes to a constant $h_0$, $H\to h_0$, in the
limit of
$t\to \pm \infty$, the universe asymptotically approaches to deSitter phase.
Therefore there does not appear Big Rip nor Big Bang singularity.

Hence, we presented several examples of inhomogeneous EOS for ideal fluid and
demonstrated how the final state of the universe filled with such fluid changes
if compare with homogeneous case. The  ideal fluid with implicit EOS may be
used
to construct the cosmologies which cross the phantom divide.

The interesting remark is in order (see also Appendix).
In principle, the more general EOS may contain the derivatives of $H$, like
$\dot H$, $\ddot H$, ... Then more general EOS than (\ref{pH11}) has the
following form:
\be
\label{dH1}
F\left(p,\rho,H,\dot H,\ddot H,\cdots\right)=0\ .
\ee
Trivial example is that
\be
\label{dH2}
p=w\rho - \frac{2}{\kappa^2}\dot H - \frac{3(1+w)}{\kappa^2}H^2\ .
\ee
By using the first (\ref{pH3}) or second (\ref{pH20}) FRW equations, we find
\be
\label{dH3}
\rho=\frac{3}{\kappa^2}H^2\ ,\quad p=-\frac{2}{\kappa^2}\dot H
   - \frac{3}{\kappa^2}H^2\ .
\ee
Therefore Eq.(\ref{dH2}) becomes an identity, which means that any cosmology
can be a solution
if  EOS (\ref{dH2}) is assumed.

Another, non-trivial example is
\be
\label{dH4}
p=w\rho - G_0 - \frac{2}{\kappa^2}\dot H + G_1{\dot H}^2\ .
\ee
Here it is supposed $G_0(1+w)>0$. If $G_1(1+w)>0$, there appears
a solution which describes an oscillating universe,
\be
\label{dH5}
H=h_0 \cos \omega t\ ,\quad a=a_0\e^{\frac{h_0}{\omega}\sin\omega t}\ .
\ee
Here
\be
\label{dH6}
h_0\equiv \kappa\sqrt{\frac{G_0}{3(1+w)}}\ ,\quad
\omega=\sqrt{\frac{3(1+w)}{G_1\kappa^2}}\ .
\ee
In case $G_1(1+w)<0$, another cosmological solution appears
\be
\label{dH7}
H=h_0\cosh \tilde\omega t\ ,\quad a=a_0\e^{\frac{h_0}{\omega}
\sinh\tilde\omega t}\ .
\ee
Here $h_0$ is defined by (\ref{dH6}) again and $\tilde\omega$ is defined by
\be
\label{dH8}
\tilde\omega=\sqrt{-\frac{3(1+w)}{G_1\kappa^2}}\ .
\ee
One can go further and present many more examples of inhomogeneous EOS
cosmology.

\section{FRW cosmology with inhomogeneous interacting fluids}

In the present section, we study FRW universe filled with two interacting
  fluids. Note that there is some interest to study the
cosmology
with homogeneous interacting fluids \cite{Q,NOT}. The inhomogeneous terms for
such cosmology may be again motivated by (bulk) viscosity account \cite{MG}.

Let us consider a system with two fluids, which satisfy the following EOS:
\be
\label{CpH1}
p_{1,2}=-\rho_{1,2} - f_{1,2}\left(\rho_{1,2}\right) - G_{1,2}\left(H\right)\ .
\ee
For simplicity,  the only case is considered that
\be
\label{CpH2}
p_\pm=w_\pm \rho_\pm - G_\pm\left(H\right)\ .
\ee
In the above equation and in the following,  the indexes $\pm$ instead of
$1,2$, as $p_{1,2}=p_\pm$ are used.
In a spatially flat FRW
universe with a scale factor $a$, the cosmological equations
are given by
\bea
\label{rhoeq}
& & \dot{\rho}_\pm+3H(\rho_\pm+p_\pm)= \mp Q\ ,\\
\label{Heq}
& & \dot{H}=-\frac{\kappa^2}{2}(\rho_++p_++\rho_-+p_-)\ , \\
\label{FRWH}
&& H^2=\frac{\kappa^2}{3}(\rho_++\rho_-)\ .
\eea
Not all of the above equations are  independent, for example, Eqs.(\ref{rhoeq})
and
(\ref{FRWH})
lead to (\ref{Heq}). From Eqs.(\ref{FRWH}), (\ref{Heq}), and the equation for
$\rho_+$ and $p_+$
of (\ref{rhoeq}), one obtains the equation for $\rho_-$ and $p_-$ of
(\ref{rhoeq}).
In \cite{NOT}, the following case has been considered:
\be
\label{NOT1}
G_\pm (H)=0\ ,\quad Q=\delta H^2 \ ,\quad w_+=0\ ,\quad w_-=-2
\ee
where $\delta$ is a constant. Then combining Eq.~(\ref{FRWH}) with
Eqs.~(\ref{rhoeq}), the explicit
     solution follows
\bea
\label{EOS40}
H &=& \frac{2}{3}\left(\frac{1}{t} + \frac{1}{t_s - t}\right)\ , \\
\label{EOS40a2}
\rho_+ &=& \frac{4}{3\kappa^2}\left(\frac{1}{t} + \frac{1}{t_s
   - t}\right)\frac{1}{t}\ , \\
\label{EOS40a}
\rho_- &=& \frac{4}{3\kappa^2}\left(\frac{1}{t} + \frac{1}{t_s
   - t}\right)\frac{1}{t_s - t} \,,
\eea
where
\be
\label{EOS40b}
t_s\equiv \frac{9}{\delta \kappa^2}\,.
\ee
In (\ref{EOS40}), it is assumed $0<t<t_s$.
The Hubble rate $H$ diverges in a finite time ($t \to t_s$)
as in the Big Rip singularity.
Therefore $t_s$ corresponds to the life time of the universe.
When $t \ll t_s$, $H$ behaves as $2/3t$, which means that the effective EOS
is given by $w_{\rm eff} \sim 0 >-1$.
On the other hand, when $t \sim t_s$, it appears $w_{\rm eff}=-2<-1$.
Therefore the crossing of phantom divide $w_{\rm eff}=-1$ occurs.

  From (\ref{CpH2} - \ref{FRWH}), we obtain
\bea
\label{CpH3}
\rho_\pm &=& \frac{3}{2\kappa^2}H^2 \nn
&& \pm \left. \frac{1}{w_+ - w_-}\right\{
G_+(H) + G_-(H) \nn
&& \left. - \frac{3}{\kappa^2}\left(1 + \frac{w_+ + w_-}{2}\right)H^2
      - \frac{2}{\kappa^2}\dot H\right\}\ ,\\
\label{CpH4}
Q&=& -\left. \frac{1}{w_+ - w_-}\right\{ \left(G_+'(H) + G_-'(H)\right)\dot H
\nn
&& \left. - \frac{6}{\kappa^2}\left(1 + \frac{w_+ + w_-}{2}\right)H \dot H
      - \frac{2}{\kappa^2}\ddot H\right\} \nn
&& + 3H\left(1+\frac{w_+ + w_-}{2}\right) \nn
&& \times \left. \frac{1}{w_+ - w_-} \right\{ G_+(H) + G_-(H) \nn
&& \left. - \frac{3}{\kappa^2}\left(1 + \frac{w_+ + w_-}{2}\right)H^2
      - \frac{2}{\kappa^2}\dot H\right\} \ \nn
&& - \frac{9\left(w_+ - w_-\right)}{4\kappa^2}H^3 \nn
&& - \frac{3}{2}H \left( G_+ (H) - G_-(H)\right) \ .
\eea
First, the case is considered that the Hubble rate $H$ satisfies the following
equation:
\be
\label{CpH5}
\dot H = S(H)\ ,
\ee
where $S(H)$ is a proper function of $H$. Hence, $Q$ can be presented as a
function
of $H$ as
\bea
\label{CpH6}
Q&=&Q(H) \nn
&=& - \left. \frac{1}{w_+ - w_-}\right\{\left(G_+'(H) + G_-'(H)\right)S(H) \nn
&& \left. + 3\left(1 + \frac{w_+ + w_-}{2}\right)H\left(G_+(H)
+ G_-(H)\right)\right\} \nn
&& + \frac{12}{\kappa^2\left(w_+ - w_-\right)}\left(1 + \frac{w_+
+ w_-}{2}\right)HS(H) \nn
&& - \frac{9}{\kappa^2\left(w_+ - w_-\right)}\nn
&& \times \left\{\left(1 + \frac{w_1 + w_2}{2}\right)^2
+ \frac{\left(w_+ - w_-\right)^2}{4}\right\}H^3 \nn
&& + \frac{2}{\kappa^2\left(w_+ - w_-\right)} S'(H) S(H) \nn
&& - \frac{3}{2}H \left(G_+ (H) - G_-(H)\right) \ .
\eea
If $Q$ is given by (\ref{CpH6}) for proper $G_p (H)$ and $S(H)$, the
solution of Eqs.
(\ref{CpH2} - \ref{FRWH}) can be obtained by solving Eq.(\ref{CpH5}) with
respect
to $H$. Then
from (\ref{CpH3}), one finds the behavior of $\rho_\pm$.
As an example, if we consider $S(H)$ given by
\be
\label{CpH7}
S(H)= - \frac{1}{h_1}\left(H - h_0\right)\ ,
\ee
the solution of (\ref{CpH6}) is given by
\be
\label{CpH8}
H=h_0 + \frac{h_1}{t - t_0}\ ,
\ee
Here $t_0$ is a constant of the integration. In the solution (\ref{CpH8}), as
$H$ behaves as
$H\sim \frac{h_1}{t - t_0}$ when $t-t_0\sim 0$, the effective $w_{\rm eff}$ is
given by
$w=-1 + \frac{2}{3h_1}$. On the other hand, as $H$ becomes a constant $h_0$
when $t$ is large,
we obtain the effective $w_{\rm eff}=1$.

Next  the simpler case is considered:
\be
\label{CpH9}
w_\pm = - 1 \pm w\ ,\quad G_\pm (H) = \pm G(H)\ .
\ee
Then (\ref{CpH3}) and (\ref{CpH4}) have the following forms:
\bea
\label{CpH10}
\rho_\pm &=& \frac{3}{2\kappa^2}H^2 \mp \frac{1}{\kappa^2w}\dot H\ ,\\
\label{CpH11}
Q&=& \frac{1}{\kappa^2 w}\ddot H - \frac{9w}{2\kappa^2}H^3 -3 HG(H)\ .
\eea
Thus, for example, for an arbitrary $G(H)$, if $Q$ is given by a function of
$H$
as
\be
\label{CpH11b}
Q=\frac{\omega^2}{\kappa^2 w} \left(H - h_0\right) - \frac{9w}{2\kappa^2}H^3 -3
HG(H)\ ,
\ee
that is,
\be
\label{CpH12}
\ddot H = \omega^2 \left(h_0 - H\right)\ ,
\ee
the solution of Eqs.(\ref{CpH2} - \ref{FRWH}) is given by
\bea
\label{CpH13}
H&=&h_0 + h_1 \sin \left(\omega t + \alpha\right)\ ,\nn
\rho_\pm &=& \frac{3}{2\kappa^2}\left(h_0 + h_1 \sin \left(\omega t +
\alpha\right)\right)^2 \nn
&& \mp \frac{h_1\omega}{\kappa^2 w} \cos \left(\omega t + \alpha\right)\ .
\eea
Here $h_1$ and $\alpha$ are constants of the integration. This
demonstrates how the inhomogeneous term modifies late-time cosmology.

Choosing $G_\pm(H)$ and $Q$, one may realize a rather general cosmology.
As was shown, if we introduce two  fluids, even without assuming
the
non-linear EOS as in (\ref{pH11}), the model crossing $w=-1$ effectively can be 
realized.
In fact, from (\ref{CpH13}) one has
\be
\label{CpH14}
\dot H = h_1 \omega \cos \left(\omega t + \alpha\right)\ ,
\ee
which changes its sign depending on time. When $\dot H>0$,
effectively $w<-1$,
and when $\dot H<0$, $w>-1$.
Note that, as a special case in (\ref{CpH11b}), we may choose,
\be
\label{CpH15}
G(H)=\frac{\omega^2}{3\kappa^2 w} \left(1 - \frac{h_0}{H}\right) - 
\frac{3w}{2\kappa^2}H^2\ ,
\ee
which gives $Q=0$. As $Q=0$, from (\ref{rhoeq}), there is no direct interaction 
between two fluids.
As is clear from (\ref{CpH13}), however, there is an oscillation in the
energy densities, which may
indicate that there is a transfer of the energy between the fluids. Hence,
the $G(H)$ term might generate
indirect transfer between two fluids.

\section{Crossing the phantom barrier with inhomogeneous EOS and 
thermodynamical
considerations}

Let us start from the EOS (\ref{EOS1}).
Assuming that $w$ crosses $-1$, which corresponds to $f(\rho)=0$, in order
that the integrations
in (\ref{EOS4}) and (\ref{tint}) are finite, $f(\rho)$ should behave as
\be
\label{Stf6}
f(\rho)\sim f_0\left(\rho - \rho_0\right)^s\ ,\quad 0<s<1\ .
\ee
Here $f(\rho_0)=0$. Since $0<s<1$, $f(\rho)$ could be
multi-valued at $\rho=\rho_0$,
in general.
Near $\rho=\rho_0$, Eq.(\ref{tint}) gives,
\be
\label{Stf7}
t-t_0\sim \frac{\left(\rho -\rho_0\right)^{1-s}}{\kappa\sqrt{3}
\rho_0f_0(1-s)}\ .
\ee
Here  $t=t_0$ when $\rho=\rho_0$.
Since
\be
\label{Stf8}
\dot H=\frac{\kappa^2}{2}f(\rho)\ ,
\ee
from the second FRW Eq.(\ref{pH20}), one finds
\bea
\label{Stf10}
\dot H &\sim& \frac{\kappa}{2^2}f_0\left(\frac{t-t_0}{t_1}
\right)^{s/(1-s)}\ ,\nn
t_1&\equiv& \frac{1}{\kappa\sqrt{3}\rho_0 f_0 (1-s)}\ .
\eea
Hence, when $s/(1-s)$ is positive odd integer, the sign of $\dot H$ changes at
$t=t_0$, which shows the
crossing $w=-1$.

In recent paper\cite{Stf}, based on consideration of mixture of two fluids:
effective quintessence and effective phantom,
the following, quite interesting EOS has been suggested:
\be
\label{Stf11}
A\rho^m + B p^m = \left(C\rho^m + D p^m\right)^\alpha \ .
\ee
Here $A$, $B$, $C$, $D$, and $\alpha$ are constants and $m$ is an integer.
This EOS can be regarded as a special case of (\ref{pH11}).
By writing $p$ as
\be
\label{Stf12}
p=Q(\rho)\rho\ ,
\ee
one obtains
\be
\label{Stf13}
\rho^{m(\alpha - 1)}=F\left(Q^m\right) \equiv \left(A + B Q^m\right)
\left( C + D Q^m \right)^{-\alpha}\ .
\ee
Since
\bea
\label{Stf14}
F'\left(Q^m\right)&=&\left(C+DQ^m\right)^{-\alpha - 1} \nn
&& \times \left(BC - \alpha AD
+ (1-\alpha)BD Q^m\right)\ ,
\eea
it follows $F'\left(Q^m\right)=0$ when
\be
\label{Stf15}
Q^m = - \frac{\frac{C}{D} - \alpha \frac{A}{B}}{1-\alpha}\ .
\ee
By properly choosing the parameters, we assume
\be
\label{Stf16}
\frac{\frac{C}{D} - \alpha \frac{A}{B}}{1-\alpha}=1\ .
\ee
When $Q\sim -1$, \be
\label{Stf17}
F(Q^m)\sim q_0 + q_2 \left(Q + 1\right)^2\ .
\ee
Here
\bea
\label{Stf18}
q_0 &=& F(1)=(A-B)(C+D)^{-\alpha} \ ,\nn
q_2 &=& \frac{1}{2}\left.\frac{d^2 F}{dQ^2}\right|_{Q=-1} \nn
&=& -\alpha (\alpha -1) (C-D)^{-\alpha -2} D^2(A-B)m^2\ .
\eea
In (\ref{Stf18}), it is supposed $m$ is an odd integer.
Solving (\ref{Stf13}) with (\ref{Stf17}) with respect to $Q$, one arrives at
\be
\label{Stf19}
Q=-1 \pm \left\{\frac{m(\alpha-1)\rho_0^{m\alpha - m -1} \left(\rho
   - \rho_0\right)}{q_2}\right\}^{1/2}\ .
\ee
Here $\rho_0$ is defined by
\be
\label{Stf19b}
q_0 = \rho_0^{m(\alpha -1)}\ .
\ee
Using (\ref{Stf12}), the function Q is
\be
\label{Stf20}
p\sim -\rho \pm \rho_0\left\{\frac{m(\alpha-1)\rho_0^{m\alpha - m -1}
\left(\rho - \rho_0\right)}{q_2}\right\}^{1/2}\ .
\ee
Comparing (\ref{Stf20})  with (\ref{Stf6}), we find
that the EOS (\ref{Stf11})
surely corresponds to $s=1/2$ case in (\ref{Stf6}).

For the EOS (\ref{Stf11}), there are interesting, exactly solvable cases. We
now consider
such a case and see that there are really the cases of EOS  crossing
barrier $w=-1$.
The energy conservation law (\ref{ppH1}) may be rewritten as follows:
\be
\label{Stf23}
p=-\rho - V\frac{d\rho}{dV}\ ,\quad V\equiv V_0 a^3\ .
\ee
Here $V_0$ is a constant with the dimension of the volume. Use of
Eq.(\ref{Stf12}) gives
\be
\label{Stf24}
0=V\frac{d\rho}{dV} + \left(1 + Q(\rho)\right)\rho\ .
\ee
Using (\ref{Stf13}), we further rewrite (\ref{Stf24}) as an equation
with respect to $Q$:
\bea
\label{Stf25}
0&=&-\frac{\left(BC - \alpha AD + BD
\left(1-\alpha\right)Q^m\right)Q^{m-1}}{(1-\alpha)
\left(C+DQ^m\right)\left(A+BQ^m\right)}V\frac{dQ}{dV} \nn
&& + 1 + Q\ .
\eea
Assuming Eq.(\ref{Stf16}), the above Eq.(\ref{Stf25}) takes a simple
form:
\be
\label{Stf26}
0=-\frac{BD\left(1 + Q^m\right)Q^{m-1}}{\left(C+DQ^m\right)
\left(A+BQ^m\right)}V\frac{dQ}{dV} + 1 + Q\ .
\ee
Especially in the simplest case $m=1$, one can easily solve (\ref{Stf26})
\be
\label{Stf27}
Q=-\frac{C - A\left(\frac{V}{V_1}\right)^\beta}{D
   - B\left(\frac{V}{V_1}\right)^\beta}\ .
\ee
Here $V_1$ is a constant of the integration and
\be
\label{Stf28}
\beta\equiv \frac{BD}{AD - BC}=\frac{1}{(1-\alpha)
\left(\frac{A}{B} - 1\right)}\ .
\ee
In the above equation, Eq.(\ref{Stf16}) is used.
Hence, when $\left(V/V_1\right)^\beta\to 0$, it follows $w=p/\rho=Q\to -
C/D$.
On the other hand, when $\left(V/V_1\right)^\beta\to \infty$, one arrives
at
$w=Q\to - A/B$.
Hence, the value of $w$ changes depending on the size of the universe.
Especially when
\be
\label{Stf29}
\frac{V}{V_1}=\left(\frac{C-D}{A-B}\right)^{1/\beta}\ ,
\ee
there occurs the crossing of phantom divide $w=Q=-1$ (compare with
\cite{Stf}).

As the inhomogeneous generalization of the EOS (\ref{Stf11}), we may
consider
\be
\label{Stf21}
A\left(\frac{3}{\kappa^2}H^2\right)^m + B p^m = \left(C\rho^m + D
p^m\right)^\alpha \ ,
\ee
or
\be
\label{Stf22}
A\rho^m + B p^m = \left(C\left(\frac{3}{\kappa^2}H^2\right)^m + D
p^m\right)^\alpha \ ,
\ee
or, more general EOS
\bea
\label{Stf30}
(A-A')\rho^m + A'\left(\frac{3}{\kappa^2}H^2\right)^m + B p^m && \nn
= \left((C-C')\rho^m + C'\left(\frac{3}{\kappa^2}H^2\right)^m + D
p^m\right)^\alpha \ . &&
\eea
By using the first FRW equation (\ref{pH3}), it folows that
the EOS (\ref{Stf21}), (\ref{Stf22}), and (\ref{Stf30})  are equivalent to
(\ref{Stf11}).
Especially if $m=1$ and (\ref{Stf16}) could be satisfied, one obtains the
solution (\ref{Stf27}).

Hence, using the first and second FRW Eqs.(\ref{pH3}) and
(\ref{pH20}),
the EOS (\ref{Stf11}) with $m=1$ can be rewritten as
\bea
\label{StfH1}
&& \frac{d^2}{dt^2}\left(a^{\frac{3}{2}\left(1-\frac{A}{B}\right)} \right) \nn
&& = 
\frac{3\kappa^2(A-B)}{4B^2}\left(\frac{3\kappa^2(C-D)}{4D^2}\right)^{-\alpha} 
\nn
&& \quad \times a^{\frac{3}{2}\left\{1 - \frac{A}{B} - \alpha\left(1 - 
\frac{C}{D}\right)\right\}}
\left\{\frac{d^2}{dt^2}\left(a^{\frac{3}{2}\left(1-\frac{C}{D}\right)}
\right)\right\}\ .
\eea
When (\ref{Stf16}) is satisfied, this second order differential Eq. looks
as
\bea
\label{StfH2}
\frac{d^2 X}{dt^2}&=&\left(\frac{4B}{3\kappa^2(A-B)}
\right)^{\alpha - 1}\alpha^\alpha
\left(\frac{d^2 X^{\frac{1}{\alpha}}}{dt^2}\right)^\alpha\ ,\nn
X&\equiv& a^{\frac{3}{2}\left(1-\frac{A}{B}\right)}\ ,
\eea
which also admits, besides the solution crossing $w=-1$ (\ref{Stf27}), a flat
universe solution
\be
\label{StfH3}
a=a_0\ ,\quad (a_0:\mbox{constant})\ ,
\ee
and deSitter universe solution
\be
\label{StfH4}
a=a_0\e^{\frac{2}{\kappa}\sqrt{\frac{B}{A-B}}\alpha
^{\frac{\alpha}{2(1-\alpha)}} t}\ .
\ee

As next generalization of (\ref{Stf11}), one may consider the following
EOS:
\bea
\label{StfH5}
&& A\rho + Bp - \frac{A-B}{\kappa^2}H^2 \nn
&& = \left( C\rho + Dp - \frac{C-D}{\kappa^2}H^2 \right)^{\alpha(H)}\ .
\eea
Here  $\alpha$ is assumed to be a function of $H$. Then
by using the first and second FRW Eqs.(\ref{pH3}) and
(\ref{pH20}), the
EOS (\ref{StfH5}) can be
rewritten as
\be
\label{StfH6}
-\frac{2B}{\kappa^2}\dot H = \left(-\frac{2D}{\kappa^2}
\dot H\right)^{\alpha(H)}\ ,
\ee
which gives
\be
\label{StfH7}
-\frac{\kappa^2}{2D}t = \int^H dH \e^{-\frac{\ln\frac{B}{D}}{\alpha(t) - 1}}\ .
\ee
As an example, for the solution (\ref{CpH13})
\be
\label{StfH8}
\omega t= \frac{1}{h_1}\int^H
\frac{dH}{\sqrt{1- \left(\frac{H-h_0}{h_1}\right)^2}}\ .
\ee
Comparing (\ref{StfH7}) with (\ref{StfH8}), in case that
\be
\label{StfH9}
h_1\omega = - \frac{\kappa^2}{2D}\ ,\quad \alpha(H)
= 1 + \frac{2\ln \frac{B}{D}}{\ln \left(1 - \left(
\frac{H-h_0}{h_1}\right)^2\right)}\ ,
\ee
the solution (\ref{CpH13}) follows from the EOS (\ref{StfH5}).

As another generalization of (\ref{Stf11}), we may consider the following EOS:
\be
\label{StfH10}
A\rho^m + Bp^m = G(H)\left( C\rho^m + Dp^m \right)^\alpha\ .
\ee
Here $G(H)$ is a function of the Hubble rate.
For simplicity,  the following case is considered
\be
\label{StfH11}
m=1\ ,\quad G(H)=\left(\frac{3}{\kappa^2}H^2 \right)^\gamma\ .
\ee
Then, writing $p$ as (\ref{Stf12}) and using $Q$, the energy looks like
\be
\label{StfH12}
\rho = (A+BQ)^\frac{1}{\gamma + \alpha -1} (C+DQ)^{-\frac{\alpha}{\gamma
+ \alpha -1}}\ ,
\ee
which corresponds to (\ref{Stf13}).
Assuming Eq.(\ref{Stf16}), by using (\ref{Stf12}),
instead of  (\ref{Stf26}), one gets
\be
\label{StfH13}
0=-\frac{(1-\alpha)BD}{(1-\alpha - \gamma)\left(C+DQ\right)
\left(A+BQ\right)}V\frac{dQ}{dV} + 1\ ,
\ee
which can be solved as
\be
\label{StfH14}
Q=-\frac{C - A\left(\frac{V}{V_1}\right)^{\tilde\beta}}{D
   - B\left(\frac{V}{V_1}\right)^{\tilde\beta}}\ .
\ee
Here $V_1$ is again a constant of the integration and
\be
\label{StfH15}
\tilde \beta\equiv \frac{(1-\alpha)BD}{(1-\alpha -\gamma)AD - BC}\ .
\ee
Then as in (\ref{Stf27}), when $\left(V/V_1\right)^\beta\to 0$, we have
$w=p/\rho=Q\to - C/D$
and when $\left(V/V_1\right)^\beta\to \infty$, we have $w=Q\to - A/B$. The
power of $V$, however, is
changed in Eq.(\ref{StfH14}) if compare with Eq.(\ref{Stf27}).
Thus, we presented number of FRW cosmologies (including oscillating
universes) filled by cosmic fluid
with inhomogeneous EOS where
phantom divide is
crossing. Definitely, one can suggest more examples or try to fit
the astrophysical data with more precise model of above sort.

In \cite{BNOV}, the thermodynamical models of the dark energy have been
constructed.
Especially it has been shown that, for the fluid with constant $w$, the free
energy
$F(T,V)$ is generally given by
\be
\label{StfT1}
F(T,V)=T\hat F\left( (T/T_0)^{1/w}(V/V_0) \right)\ .
\ee
Here $T$ is the temperature and $V$ is the volume of the universe. For the
dimensional reasons,
the positive parameters $T_0$ and $V_0$ are introduced.

The interesting question is what happens with the entropy
when the value of $w$ crosses $-1$.
As a model,  the case that $w=Q$ depends on $V$ as in (\ref{Stf27}) may be
considered:
\be
\label{StfT2}
w=w(V)=\frac{w_0 + w_1\left(\frac{V}{V_0}\right)^\beta}{1
+ \left(\frac{V}{V_0}\right)^\beta}\ .
\ee
When $\beta>0$,  $w\to w_0$ for small universe and $w\to w_1$ for large
universe.

The specific dependence of free energy may be taken as below
\be
\label{StfT3}
F=\frac{f_0 T}{T_0}
\left\{\left(\frac{T}{T_0}\right)^{\frac{1}{w(V)}}\frac{V}{V_0}
\right\}^{\gamma}\ .
\ee
Here $\gamma$ is a constant. When $\gamma=1$ and $w$ is a constant,
  the free energy is proportional to the
volume. For usual matter, due to self-interaction and related effects,
$\gamma$ is not always unity.
Then, the pressure $p$, the energy density $\rho$, and the entropy
${\cal S}$ are given by
\bea
\label{StfT4}
p &=& - \frac{\partial F}{\partial V} \nn
&=& - \frac{f_0\gamma}{V_0}\left(\frac{T}{T_0}\right)^{1 + \frac{\gamma}{w(V)}}
\left(\frac{V}{V_0}\right)^{\gamma -1} \nn
&& \times \left\{ 1 +
\gamma \ln \left(\frac{T}{T_0}\right)
\frac{\left(w_1 - w_0\right)\beta \left(\frac{V}{V_0}\right)^\beta}
{\left( w_0 + w_1\left(\frac{V}{V_0}\right)^\beta\right)^2}\right\}\ ,\nn
\rho &=& \frac{1}{V}\left(F - T \frac{\partial F}{\partial T}\right) \nn
&=& - \frac{f_0\gamma}{w V_0}\left(\frac{T}{T_0}\right)^{1 + 
\frac{\gamma}{w(V)}}
\left(\frac{V}{V_0}\right)^{\gamma -1}\ , \nn
{\cal S} &=& - \frac{\partial F}{\partial T} \nn
&=& - \frac{f_0}{T_0}\left(1
+ \frac{\gamma}{w}\right)\left(\frac{T}{T_0}\right)^{\frac{\gamma}{w(V)}}
\left(\frac{V}{V_0}\right)^{\gamma}\ .
\eea
In the pressure $p$, the second term in large $\{\ \}$ comes from $V$
dependence of $w$
in (\ref{StfT2}), which vanishes for large or small universe ($V\to \infty$ or
$V\to 0$).
Hence, for small or large universe $p/\rho\to w(V) \to w_{0,1}$.
As seen
from the expression for
${\cal S}$, the sign of the entropy changes at
\be
\label{StfT5}
w=-\gamma\ .
\ee
If $\gamma=1$, the sign of the entropy ${\cal S}$ changes when crossing
$w=-1$ (the entropy becomes negative when $w$ is less than $-1$ as it was
observed in \cite{BNOV}), but in the case that
\be
\label{StfT6}
\gamma <|w_0|,\,|w_1|\ ,
\ee
the entropy does not change its sign.

We should note that the expressions  (\ref{StfT4}) are not well-defined,
unless $\gamma=0$, when $w=0$, which corresponds to dust.
One may assume $0<\gamma<w_0\ll 1$ and $w_1\lesssim -1$.
Then as clear from (\ref{StfT2}), $w$ changes from $w_0\sim 0$ for small 
universe to $w_1\lesssim -1$
for large universe and crosses $-1$.
Since we always have $\left| \gamma/w_0 \right| < 1$ and
  therefore $1+\gamma/w>0$, the entropy ${\cal S}$
(\ref{StfT4}) is always positive and does not change its sign
as long as $f_0<0$. This explicitly demonstrates very beatiful phenomenon:
there exist thermodynamical models for dark energy with crossing of
phantom divide. Despite the preliminary expectations, the entropy of such
dark energy universe even in its phantom phase may be positive!

\section{Discussion}

In summary, the effect of modification of general EOS of dark energy
ideal fluid by the insertion of inhomogeneous, Hubble parameter dependent
term in the
late-time universe is considered. Several explicit examples of such term
  which is motivated by time-dependent bulk viscosity or deviations from
general relativity
are considered. The corresponding late-time FRW cosmology (mainly, in its
phantom epoch) is described. It is demonstrated how the structure of
future singularity is changed thanks to generalization of dark energy EOS.
The number of FRW cosmologies admitting the crossing of phantom barrier
are presented. The inhomogeneous term in EOS helps to realize such a
transition in a more natural way.

It is interesting that in the case when universe is filled with two
  interacting fluids (for instance, dark energy and dark matter) the Hubble
parameter dependent term may effectively absorb the coupling between the
fluids. Again, in case of two dark fluids the phantom epoch with possibility of 
crossing
of $w=-1$ barrier occurs is constructed.
It is also very interesting that there exists thermodynamical dark energy model 
where despite the preliminary expectations\cite{BNOV} the entropy
in phantom epoch may be positive. This is caused by crossing of phantom barrier.

As it was demonstrated making the dark energy EOS more general, this extra 
freedom in inhomogeneous term brings a number of new possibilities to
construct the late-time universe. One can go even further , assuming
that inhomogeneous terms in EOS are not restricted by energy conservation law 
(as it is often the case in braneworld approach). Nevertheless, only
more precise astrophysical data will help to understand which of number of EOS 
of the universe under consideration (in other words, dark energy models) is 
realistic.

\section*{ACKNOWLEDGEMENTS}

We thank S. Tsujikawa for participation at the
early stage of this work. The research by SDO has been partially supported
by RFBR grant 03-01-00105 and LRSS grant 1252.2003.2.

\appendix

\section{Inhomogeneous terms from modified gravity}

Let us consider the possibility to obtain the  inhomogeneous EOS
from the modified  gravity.
As an illustrative example,  the following action is considered:
\be
\label{HD1}
S=\int d^4 x \sqrt{-g}\left(\frac{1}{2\kappa^2} + {\cal L}_{\rm matter}
+ f(R)\right)\ .
\ee
Here $f(R)$ can be an arbitrary function of the scalar curvature $R$ and
${\cal L}_{\rm matter}$ is the Lagrangian for the matter.
In the FRW universe, the gravitational
equations are:
\bea
\label{HD2}
0&=& - \frac{3}{\kappa^2}H^2 + \rho
- f\left(R=6\dot H + 12 H^2\right) \nn
&& + 6\left(\dot H + H^2 - H \frac{d}{dt}\right) \nn
&& \times f'\left(R=6\dot H + 12 H^2\right) \ ,\\
\label{HD3}
0&=& \frac{1}{\kappa^2}\left(2\dot H + 3H^2\right) + p + f\left(R=6\dot H
+ 12 H^2\right) \nn
&& + 2\left( - \dot H - 3H^2 + \frac{d^2}{dt^2} + 2H \frac{d}{dt}\right) \nn
&& \times f'\left(R=6\dot H + 12 H^2\right)\ .
\eea
Here $\rho$ and $p$ are the energy density and the pressure coming from ${\cal
L}_{\rm matter}$.
They may satisfy the equation of state like $p=w\rho$.
One may now define the effective energy density $\tilde \rho$ and $\tilde
p$ by
\bea
\label{HD4}
\tilde\rho &\equiv& \rho
   - f\left(R=6\dot H + 12 H^2\right) \nn
&& + 6\left(\dot H + H^2 - H \frac{d}{dt}\right) \nn
&& \times f'\left(R=6\dot H + 12 H^2\right) \ ,\\
\label{HD5}
\tilde p&=& p + f\left(R=6\dot H + 12 H^2\right) \nn
&& + 2\left( - \dot H - 3H^2 + \frac{d^2}{dt^2} + 2H \frac{d}{dt}\right) \nn
&& \times f'\left(R=6\dot H + 12 H^2\right)\ .
\eea
Thus, it follows
\bea
\label{HD6}
\tilde p&=& w\tilde \rho  + (1+w)f\left(R=6\dot H + 12 H^2\right) \nn
&& + 2\left( \left(-1 - 3w\right) \dot H - 3\left(1+w\right) H^2
+ \frac{d^2}{dt^2} \right.\nn
&& \left. + \left(2 + 3w\right) H \frac{d}{dt}\right)
f'\left(R=6\dot H + 12 H^2\right) \ .
\eea
In the situation where the derivative of $H$ can be neglected as
$\dot H\ll H^2$ or $\ddot H\ll H^3$,
we find
\bea
\label{HD7}
\tilde p&\sim& w\tilde \rho  + G(H)\ ,\nn
G(H)&\equiv& (1+w)f\left(R=12 H^2\right) \nn
&& - 3\left(1+w\right) H^2
f'\left(R=12 H^2\right)\ .
\eea
Typically $H$ has a form like $H\sim h_0/\left(t-t_1\right)$ or
$H\sim h_0/\left(t_2 - t\right)$, with $h_0=2/3(w+1)$, corresponding
to (\ref{ppH4}). Hence, the condition $\dot H\ll H^2$ or
$\ddot H\ll H^3$ requires $h_0\gg 1$, which shows $w\sim -1$
as in the modern universe. This supports our observation that
inhomogeneous terms may be the effective ones which are predicted
due to currently modified gravity theory.

The modification of the EOS by $G(H)$ terms might come also from the
braneworld scenario.
Indeed, the single brane model is described by the following simple action
\bea
\label{DDG1}
S&=&\frac{M_{\rm Pl}^2}{r_c}\int d^4 x dy \sqrt{-g^{(5)}} R^{(5)} \nn
&& + \int d^4x \sqrt{-g}\left( M_{\rm Pl}^2 R + {\cal L}_{\rm matter}\right)\ .
\eea
Here $M_{\rm Pl}^2=1/8\pi G$, $y$ is the coordinate of the extra dimension,
and ${\cal L}_{\rm matter}$ is the Lagrangian density of the matters on the
brane.
The five-dimensional quantities are denoted by suffix ``(5)''.
In ref.\cite{DDG} it has been shown that the FRW equation for 4d brane
universe could be
given by
\be
\label{DDG2}
\frac{3}{\kappa^2}\left(H^2 \pm \frac{H}{r_c}\right)=\rho\ .
\ee
Here $\rho$ is the matter energy density coming from ${\cal L}_{\rm matter}$.
More general case is considered in ref.\cite{DT} where the FRW equation is
modified as
\be
\label{DDG3}
\frac{3}{\kappa^2}\left(H^2 - \frac{H^\alpha}{r_c^{2-\alpha}}\right)=\rho\ .
\ee
Here $\alpha$ is a constant.
One may assume that the matter energy density $\rho$ satisfies the energy
conservation as in (\ref{ppH1}).
Then from (\ref{DDG2}), we find
\be
\label{DDG4}
   -\frac{2}{\kappa^2}\left(1 - \frac{\alpha H^{\alpha
   -2}}{2r_c^{2-\alpha}}\right)\dot H = \rho + p\ .
\ee
By comparing (\ref{DDG3}) with the first FRW equation (\ref{pH3}) and
(\ref{DDG4}) with the second
FRW equation (\ref{pH20}), one may define the effective energy density
$\tilde
\rho$ and pressure $\tilde p$
as
\be
\label{DDG5}
\tilde\rho \equiv \rho + \frac{3H^\alpha}{\kappa^2 r_c^{2-\alpha}}\ ,\quad
\tilde p \equiv - \frac{3H^\alpha}{\kappa^2 r_c^{2-\alpha}}
    - \frac{\alpha H^{\alpha -2}\dot H}{\kappa^2 r_c^{2-\alpha}}\ ,
\ee
They satisfy the first (\ref{pH3}) and second (\ref{pH20}) FRW equations:
\be
\label{DDG6}
\frac{3}{\kappa^2}H^2=\tilde \rho\ ,\quad - \frac{2}{\kappa^2}\dot H
= \tilde\rho + \tilde p\ .
\ee
If it is also assumed the matter energy density $\rho$ and the matter
pressure $p$
satisfy the EOS like
$p=w\rho$, the effective EOS for $\tilde\rho$ and $\tilde p$ is given
by
\be
\label{DDG7}
\tilde p = w \tilde\rho - (1+w) \frac{3H^\alpha}{\kappa^2 r_c^{2-\alpha}}
    - \frac{\alpha H^{\alpha -2}\dot H}{\kappa^2 r_c^{2-\alpha}}\ .
\ee
Especially if one can neglect $\dot H$, it follows
\be
\label{DDG8}
\tilde p \sim w \tilde\rho - (1+w) \frac{3H^\alpha}{\kappa^2 r_c^{2-\alpha}}
\ .
\ee
This shows that brane-world scenario may also suggest various forms of
inhomogeneous modification for effective EOS of matter on the brane.

\end{document}